\documentclass[prl,aps,floats,twocolumn]{revtex4}

\usepackage[dvips]{graphicx}
\usepackage{amssymb} 
\usepackage{amsmath}
\usepackage{ctable}

\begin{document}

\title{Non-Commutative Geometry, Non-Associative Geometry \\
and the Standard Model of Particle Physics}

\author{Latham Boyle and Shane Farnsworth} 

\affiliation{Perimeter Institute for Theoretical
  Physics, Waterloo, Ontario N2L 2Y5, Canada} 

\date{January 2014}

\begin{abstract}
  Connes' notion of non-commutative geometry (NCG) generalizes 
  Riemannian geometry and yields a striking reinterepretation of the standard model of 
  particle physics, coupled to Einstein gravity.  We suggest a simple reformulation with 
  two key mathematical advantages: (i) it unifies many of the traditional NCG 
  axioms into a single one; and (ii) it immediately generalizes from non-commutative
  to non-associative geometry.  Remarkably, it also resolves a long-standing problem 
  plaguing the NCG construction of the standard model, by precisely eliminating from the action
  the collection of 7 unwanted terms that previously had to be removed by an extra, 
  non-geometric, assumption.  With this problem solved, the NCG algorithm for constructing 
  the standard model action is tighter and more explanatory than the traditional 
  one based on effective field theory.
\end{abstract}

\maketitle

\section{Introduction}

Since the early 1980Õs, Connes and others have been developing the 
subject of non-commutative geometry (NCG) \cite{Connes:1994yd,Connes:1985}.  
Its mathematical interest stems from the fact that it provides a natural generalization
of Riemannian geometry (much as Riemannian geometry, in turn, provides a 
natural generalization of Euclidean geometry).  Its physical interest stems from the 
fact that it suggests an elegant geometric reinterpretation of the 
standard model (SM) of particle physics (coupled to Einstein gravity) \cite{Connes:1990qp, 
Connes:1996gi, Chamseddine:1991qh, Chamseddine:1996zu, Barrett:2006qq, 
Connes:2006qv, Chamseddine:2006ep, Chamseddine:2007hz, Chamseddine:2007ia}.
For an introduction, see Refs.~\cite{Chamseddine:2010ps, 
vandenDungen:2012ky}. 
Here we propose a simple reformulation of the traditional NCG
formalism that has fundamental advantages from both the mathematical
and physical standpoint.  

Our idea, in brief, is as follows.  In the traditional NCG formalism, 
a geometry is described by a so-called real spectral triple 
$\{A,H,D,J,\gamma\}$ (again, see {\it e.g.}\ Refs.~\cite{Chamseddine:2010ps, 
vandenDungen:2012ky} for an introduction).  The essence of
our reformulation is the observation that these elements naturally fuse 
to form a new algebra $B$; and that many of the traditional NCG axioms may 
then be recovered by the single requirement that $B$ is an associative
$\ast$-algebra. 

From the mathematical standpoint, this 
has two advantages.  First, it unifies many of the traditional NCG
axioms into a single one, thereby clarifying their meaning.
Second, it naturally generalizes Connes' framework from 
non-commutative to non-associative geometry.  Remarkably, 
it also has an unexpected physical consequence: it 
solves a key problem (highlighted {\it e.g.}\ in \cite{Chamseddine:2006ep}
and the concluding section of \cite{Chamseddine:2007hz}) which has
plagued the NCG construction of the SM action functional. It does so by precisely 
eliminating from the action the collection of 7 unwanted terms that previously 
had to be removed by an extra (empirically-motivated, non-geometrical) assumption 
(called the ``massless photon" assumption in \cite{Chamseddine:2007hz}),
in order to obtain agreement between the usual SM action 
and the action obtained from the NCG construction 
\cite{Chamseddine:2006ep, Chamseddine:2007hz, 
Chamseddine:2007ia, vandenDungen:2012ky}.

With this problem solved, NCG gives an 
algorithm for constructing the SM which is tighter and more 
explanatory than the traditional one based on effective field theory (EFT). In the usual EFT construction, one must give three independent inputs: (i) the symmetries of the action; (ii) the fermions and their representations; and (iii) the scalars and their representations.  In the NCG construction, one only needs two inputs: (i') the choice of algebra $A$ (which determines the symmetry of the action); and (ii') the representation of $A$ (which determines the fermions and their representations).  Note that the third EFT
input (the scalars and their representations), is not required in the NCG approach: {\it i.e.}\ the number of higgs bosons and their representations are a predicted {\it output}, once the symmetries and the fermionic representations are specified! Furthermore, the NCG construction explains various aspects of the SM fermionic representations which are unexplained by the usual EFT construction (such as why all SM fermions transform in either the trivial or fundamental representation of the gauge groups $SU(2)$ and $SU(3)$ -- see \cite{Krajewski:1996se, Chamseddine:2007hz,  Chamseddine:2007ia} for
more details).  This striking situation is reviewed at greater length in \cite{BoyleFarnsConference}.

\section{Reformulation of Connes' Framework}

In this section, we present our formulation in two steps. In the first 
step, we explain how to extend $A$ (a $\ast$-algebra) to $\Omega A$ 
(the differential graded $\ast$-algebra of forms over $A$), 
even when $A$ is non-associative.  In the second step, we explain 
how $H$ may be promoted to a bimodule over $A$
by defining a new algebra $B_{0}=A\oplus H$; and similarly,
$H$ may be promoted to a bimodule over $\Omega A$ by 
defining a new algebra $B=\Omega A\oplus H$.  In the 
process, we obtain a new view of the operator $J$, and 
Connes' ``order-zero" and ``order-one" axioms.  Again, a
key virtue of our reformulation is that it naturally extends to the 
case where $A$ is non-associative.  (The mathematical and
physical motivations for generalizing non-commutative geometry 
to non-associative geometry
are explained in \cite{Farnsworth:2013nza}; for earlier work 
related to non-associative geometry, see \cite{Wulkenhaar:1996xk, 
Wulkenhaar:1996at, Wulkenhaar:1996av, Wulkenhaar:1996pv, AkramiMajid}.)

\paragraph{\textbf{Step 1: Promoting $A\to\Omega A$.}}

Let $A$ be a unital $\ast$-algebra over a field $F$.  ($A$ may be 
non-commutative, or even non-associative.)  We introduce,
for each element $a\in A$, a corresponding formal symbol
$\delta[a]$.  $\Omega A$ is the algebra generated by $A$ and
these differentials $\delta[a]$, modulo the relations  $\delta[fa]=f\delta[a]$, 
$\delta[a+a']=\delta[a]+\delta[a']$, $\delta[aa']=\delta[a]a'+
a\delta[a']$ (with $f\in F$, and $a,a'\in A$) and, in addition,
modulo appropriate associativity relations.  For example, in 
the usual case where $A$ is associative, we take $\Omega A$ 
to be associative as well, and impose relations like 
$(a\delta[b])c=a(\delta[b]c)$, etc; in this way, we recover
the usual algebra $\Omega A$ defined, {\it e.g.}\, in Section 6.1 of 
Ref.~\cite{Landi:1997sh}.  More generally, when $A$ 
is non-associative, $\Omega A$ must be equipped with compatible
associativity relations (for example, if $A$ is an alternative algebra,
then $\Omega A$ could be taken to be alternative as well; other
examples are given in Ref.~\cite{NonAssocExamples}).

We can write $\Omega A=\Omega^{0}\!A\oplus
\Omega^{1}\!A\oplus\Omega^{2}\!A\oplus\ldots$ where
$\Omega^{m}\!A$ is the subspace of $\Omega A$
consisting of linear combinations of terms containing $m$ 
differentials $\delta[a]$: for example,
$(a_{1}\delta[a_{2}])(\delta[a_{3}]\delta[a_{4}])+
(\delta[a_{5}](\delta[a_{6}]\delta[a_{7}]))a_{8}$
is an element of $\Omega^{3}\!A$.  In particular, $\Omega^{0}\!A=A$;
and if $\omega_{m}\in\Omega^{m}\!A$ and 
$\omega_{n}\in\Omega^{n}\!A$ then $\omega_{m}
\omega_{n}\in\Omega^{m+n}\!A$, so $\Omega A$
is graded.  If we define $\delta[a^{\ast}]=-\delta[a]^{\ast}$,
then $\Omega A$ is a $\ast$-algebra, with its
$\ast$-operation naturally inherited from $A$.  Note that we 
can interpret $\delta$ as a linear map from $\Omega^{0}\!A\to
\Omega^{1}\!A$; and this may, in turn, be promoted to 
a linear map $d:\Omega^{m}\!A\to\Omega^{m+1}\!A$
which (even in the non-associative case) may be defined 
recursively by requiring it to satisfy
a graded Leibniz rule $d[\omega_{m}\omega_{n}]
=d[\omega_{m}]\omega_{n}+(-1)^{m}\omega_{m}
d[\omega_{n}]$, along with the conditions
$d[a]=\delta[a]$ and $d^{2}[a]=0$
($\omega_{m}\in\Omega^{m}\!A$, $\omega_{n}\in
\Omega^{n}\!A$, $a\in A$).  It follows that $d^{2}[\omega]=0$
($\omega\in \Omega A$), so that $(\Omega A,d)$ is a differential
graded $\ast$-algebra.

\paragraph{\textbf{Step 2: Promoting $\Omega^{0}\!A\to B_{0}$ and $\Omega A\to B$.}}

Usually, in non-commutative geometry, one starts by defining (in two 
steps) a bi-representation of $A$ on $H$, so that $H$ becomes 
a bi-module over $A$.  In the first step, one defines a left action 
of $A$ on $H$ -- {\it i.e.}\ a bilinear product $ah=L_{a}h\in H$ 
between elements $a\in A$ and $h\in H$.  In the second step 
one uses $J$, an anti-unitary operator on $H$, to define a 
corresponding right-action of $A$ on $H$ -- {\it i.e.}\ another 
bilinear product $ha\equiv R_{a}h\in H$ given by
$R_{a}\equiv J L_{a^{\ast}}J^{\ast}$.  The left and
right action are required to satisfy the so-called order-zero
condition $(ah)b=a(hb)$ or, equivalently, $[L_{a},R_{b}]=0$
($\forall a,b\in A$).

We would like to reformulate this construction in a way that makes 
sense even when $A$ is non-associative.  Fortunately, the natural 
definition of a ``bi-representation" of a non-associative algebra 
(or, equivalently, a ``bimodule" over a non-associative algebra) was
found long ago (perhaps by Samuel Eilenberg \cite{Eilenberg}), 
and is explained simply and succinctly in Ch.\ II.4 of Ref.~\cite{Schafer}.
The idea is that a bimodule $H$ over $A$ is nothing but a new algebra 
\begin{equation}
  B_{0}=A\oplus H,
\end{equation}
with the product between two elements of $B_{0}$ 
($b_{0}=a+h$ and $b_{0}'=a'+h'$) given by
\begin{equation}
  b_{0}b_{0}'=aa'+ah'+ha',
\end{equation}
where $aa'\in A$ is the product inherited from $A$, while
$ah'\in H$ and $ha'\in H$ are precisely the left- and right-actions
defined above.  In this language, two familiar
axioms of non-commutative geometry -- namely, (i) the 
associativity of $A$ and (ii) the order-zero condition -- 
are condensed into the single assumption that $B_{0}$ is an
associative algebra.  Furthermore, the familiar definition of right-action 
in terms of left action, $R_{a}=J L_{a^{\ast}} J^{\ast}$, is
reinterpreted as the statement that the map 
\begin{equation}
  \label{b0_star}
  b_{0}=a+h\quad\to\quad b_{0}^{\ast}=a^{\ast}+Jh
\end{equation}
is an anti-automorphism of $B_{0}$, with period
$2$ when the KO dimension is $0$, $1$, $6$ or $7$ 
mod $8$ ({\it i.e.}\ when $J^{2}=1$) and period $4$ 
when the K0 dimension is $2$, $3$, $4$ or $5$ mod $8$
({\it i.e.}\ when  $J^{2}=-1$).  In particular, when
$J^{2}=1$, $B_{0}$ is a $\ast$-algebra, with 
$\ast$ operation given by (\ref{b0_star}).  The advantage of 
this reformulation is that it continues to make sense when $A$ 
is non-associative: in this case, $B_{0}$ is non-associative, 
too, and the familiar order zero 
condition is replaced by a compatible restriction on 
the associativity properties of $B_{0}$.  For example, if
$A$ is an alternative algebra, like the octonions, we can
require $B_{0}$ to be an alternative algebra, too.  The interpretation
of $J$ in terms of the anti-automorphism (\ref{b0_star}) is
unaffected.

Just as we give a bi-representation of $A$ on $H$ by 
defining a new algebra $B_{0}=A \oplus H$, we 
give a bi-representation of $\Omega A$ on $H$ by 
defining a new algebra 
\begin{equation}
  B=\Omega A\oplus H
\end{equation}
with the product between two elements of $B$
($b=\omega+h$ and $b'=\omega'+h'$) given by
\begin{equation}
  bb'=\omega\omega'+\omega h'+h\omega',
\end{equation}
where $\omega\omega'\in\Omega A$ is the product
inherited from $\Omega A$, while $\omega h'\in H$
and $h\omega'\in H$ are bilinear products that
define the left-action and right-action of $\Omega A$
on $H$\footnote{For brevity, we are skipping over the
issue of junk forms, and the corresponding distinction
between $\Omega A$ and $\Omega A_{D}=\Omega A/J$
(see Sec. 6.2 of Ref.~\cite{Landi:1997sh}), since this 
nuance is not important for our present purposes; but we 
note that they may be readily encorporated in our approach.}.
Thus, just as $A=\Omega^{0}\!A$ is a 
subalgebra of $\Omega A$, $B_{0}$ is a corresponding
subalgebra of $B$.  The elements $\omega\in\Omega A$ are
linear combinations of products of $a$'s and $\delta[a]$'s:
having already introduced the left- and right-action 
of $a$ on $H$ ($ah=L_{a}h$ and $ha=R_{a}h$), we
now obtain the left- and right-action of $\delta[a]$
on $H$ by regarding $D$, a hermitian operator on $H$,
as the representation on $H$ of the map $\delta:\Omega^{0}\!A
\to\Omega^{1}\!A$, and requiring that it satisfies the corresponding
Leibniz rule: $D(ah)=\delta[a]h+a(Dh)$.
This gives $\delta[a]h=
[D,L_{a}]h$ and $h\,\delta[a]=J[D,L_{a}]^*J^*h$.  We see that
Connes' order-zero and order-one conditions are
reinterpreted here as associativity conditions on 
$B$: in particular, the order zero condition
$[L_{a},R_{b}]=0$ is the requirement that the 
associator $[\omega_{0},h,\omega_{0}']$ vanishes,
while the order one conditions $[L_{a},[D,R_{b}]]=0$ and 
$[[D,L_{a}],R_{b}]=0$ are the requirements that the 
associators $[\omega_{0},h,\omega_{1}]$ and
$[\omega_{1},h,\omega_{0}]$ vanish.  In the case where $A$ is 
non-associative, the familiar order-zero and order-one conditions 
are replaced by compatible associativity constraints on $B$.

\section{Application to the Standard Model}

In this section, we first review the traditional formulation of the 
standard model in non-commutative geometry, and then explain 
how our reformulation naturally yields a new constraint that
resolves a well-known puzzle that arises in the traditional formulation.
For clarity, we will deal in this section with a single-generation of 
standard model fermions; the extension to the full set of three 
generations is straightforward.

The standard model is described by a finite-dimensional real 
spectral triple $\{A, H, D, J, \gamma\}$
of K0 dimension 6.  $A$ is a $\ast$-algebra given by
$\mathbb{C}\oplus\mathbb{H}\oplus M_{3}(\mathbb{C})$, where
$\mathbb{C}$ is the algebra of complex numbers, $\mathbb{H}$ is
the algebra of quaternions, and $M_{3}(\mathbb{C})$ is the algebra
of $3\times 3$ complex matrices.  $H$ is a 32-dimensional
complex Hilbert space (32 is the number of fermionic degrees of 
freedom in a standard model generation, including the
right-handed neutrino).  To describe the action of $\gamma$ and 
$J$ on $H$, it is convenient to split $H$ into four 8-dimensional 
subspaces $H=H_{R}\oplus H_{L}\oplus\bar{H}_{R}\oplus\bar{H}_{L}$.  
Here $H_{R}$ and $H_{L}$ contain the right-handed and left-handed 
particles, while $\bar{H}_{R}$ and $\bar{H}_{L}$ contain the
corresponding anti-particles.  If $h_{R}\in H_{R}$ is a right-handed 
particle (with $\bar{h}_{R}\in\bar{H}_{R}$ the corresponding anti-particle)
and $h_{L}\in H_{L}$ is a left-handed particle (with 
$\bar{h}_{L}\in\bar{H}_{L}$ the corresponding anti-particle),
then the helicity operator $\gamma$ and the anti-linear charge 
conjugation operator $J$ act as follows:
\begin{equation}
  \renewcommand{\arraystretch}{1.2}
  \begin{array}{rclrclrclrcl}
    \!\!\!\!\!\gamma h_{R}\!\!&\!\!=\!\!&\!\!-h_{R},\;\;&
    \gamma h_{L}\!\!&\!\!=\!\!&\!\!h_{L},\;\;&
    \gamma \bar{h}_{R}\!\!&\!\!=\!\!&\!\!\bar{h}_{R},\;\;,&  
    \gamma \bar{h}_{L}\!\!&\!\!=\!\!&\!\!-\bar{h}_{L}, \\
    \!\!\!\!\!J h_{R}\!\!&\!\!=\!\!&\!\!\bar{h}_{R},\;\;&
    J h_{L}\!\!&\!\!=\!\!&\!\!\bar{h}_{L},\;\;&
    J \bar{h}_{R}\!\!&\!\!=\!\!&\!\!h_{R},\;\;&
    J \bar{h}_{L}\!\!&\!\!=\!\!&\!\!h_{L}.
  \end{array}
\end{equation}
To describe the action of $A$ on $H$, it is convenient 
to further split each of the four spaces $(H_{R}, H_{L}, \bar{H}_{R}, 
\bar{H}_{L})$ into a lepton and quark subspace: $H_{R}=L_{R}\oplus 
Q_{R}$, $H_{L}=L_{L}\oplus Q_{L}$, $\bar{H}_{R}=\bar{L}_{R}\oplus
\bar{Q}_{R}$, and $\bar{H}_{L}=\bar{L}_{L}\oplus\bar{Q}_{L}$.
Each of the four lepton spaces $\{L_{R}, L_{L}, \bar{L}_{R}, \bar{L}_{L}\}$ 
is a copy of $\mathbb{C}^{2}$; an element of any of these four spaces 
correspondingly carries a doublet (neutrino vs. electron) index.  Each 
of the four quark spaces $\{Q_{R}, Q_{L}, \bar{Q}_{R}, \bar{Q}_{L}\}$ is 
a copy of $\mathbb{C}^{2}\otimes \mathbb{C}^{3}$; an element of any 
one of these four spaces correspondingly carries two indices: a doublet 
(up quark vs. down quark) index and a triplet (color) index.  Now consider 
an element $a=\{\lambda,q,m\}\in{\cal A}_{F}$, where $\lambda\in\mathbb{C}$ 
is a complex number, $q\in\mathbb{H}$ is a quaternion, and 
$m\in M_{3}(\mathbb{C})$ is a $3\times 3$ complex matrix, and write
\begin{equation}
  q=\left(\begin{array}{cc} \alpha & \beta \\
  -\bar{\beta} & \bar{\alpha}\end{array}\right),\quad\qquad
  q_{\lambda}=\left(\begin{array}{cc}
  \lambda & 0 \\ 0 & \bar{\lambda}\end{array}\right),
\end{equation}
where $\alpha$ and $\beta$ are complex numbers.  Here $q$
is the standard $2\times2$ complex matrix representation of a
quaternion, and $q_{\lambda}$ is the corresponding diagonal
embedding of $\mathbb{C}$ in $\mathbb{H}$.  Then $L_{a}$ 
(the left action of $a$ on ${\cal H}$) is given by
\begin{equation}
  \label{L_a}
  \renewcommand{\arraystretch}{1.2}
  \begin{array}{lclclcl}
    L_{a}L_{R}&=&q_{\lambda}L_{R},&\quad\qquad&
    L_{a}L_{L}&=&q L_{L}, \\
    L_{a}Q_{R}&=&q_{\lambda} Q_{R},&\quad\qquad&
    L_{a}Q_{L}&=&q Q_{L}, \\
    L_{a}\bar{L}_{R}&=&\lambda\mathbb{I}_{2\times2}
    \bar{L}_{R},&\quad\qquad&
    L_{a}\bar{L}_{L}&=&\lambda\mathbb{I}_{2\times2}
    \bar{L}_{L}, \\
    L_{a}\bar{Q}_{R}&=&m \bar{Q}_{R},&\quad\qquad&
    L_{a}\bar{Q}_{L}&=&m \bar{Q}_{L}. \\
  \end{array}
\end{equation}
where $q$, $q_{\lambda}$ and $\lambda\mathbb{I}_{2\times2}$
act on the doublet index, while $m$ acts on the color index.

$D$ obeys the following four geometric constraints: $D^{\dagger}=D$, $\{D,
\gamma\}=0$, $[D,J]=0$ and $[[D,L_{a}],R_{b}]=0$.  
In the basis $\{L_{R},Q_{R},L_{L},Q_{L},
\bar{L}_{R},\bar{Q}_{R},\bar{L}_{L},\bar{Q}_{L}\}$, these imply
\begin{equation}
  \label{D}
  D=\left(
  \renewcommand{\arraystretch}{1.2}
  \begin{array}{cc|cc|cc|cc}
  \;\;0\;\;\, & \;\;0\;\; & y_{l}^{\dagger} & \;\;0\;\; & 
  m^{\dagger} & n^{\dagger} & \;\;0\;\; & \;\;0\;\; \\
  0 & 0 & 0 & y_{q}^{\dagger} & \bar{n} & 0 & 0 & 0 \\
  \hline
  y_{l}^{} & 0 & \;\;0\;\; & 0 & \;\;0\;\; & \;\;0\;\; & 0 & 0 \\
  0 & y_{q}^{} & 0 & 0 & 0 & 0 & 0 & 0 \\
  \hline
  m & n^{T} & 0 & 0 & 0 & 0 & y_{l}^{T} & 0  \\
  n & 0 & 0 & 0 & 0 & 0 & 0 & y_{q}^{T} \\
  \hline
  0 & 0 & 0 & 0 & \bar{y}_{l}^{} & 0 & 0 & 0 \\
  0 & 0 & 0 & 0 & 0 & \bar{y}_{q}^{} & 0 & 0 
\end{array}\right)
\end{equation}  
where
\begin{equation}
  y_{l}^{}=\left(\begin{array}{cc} 
  y_{l,11}^{} & y_{l,12}^{} \\ y_{l,21}^{} & y_{l,22}^{} \end{array}\right)
  \quad{\rm and}\quad
  y_{q}^{}=\left(\begin{array}{cc} 
  y_{q,11}^{} & y_{q,12}^{} \\ y_{q,21}^{} & y_{q,22}^{} \end{array}\right)
\end{equation}
are arbitrary $2\times2$ matrices that act on the doublet indices 
in the lepton and quark sectors, respectively, while 
\begin{equation}
  \label{mn}
  m=\left(\begin{array}{cc}
  a & b \\ b & 0 \end{array}\right)
  \quad{\rm and}\quad
  n=\left(\begin{array}{cc}
  \vec{c} & \vec{d} \\
  \vec{0} & \vec{0} \end{array}\right)
\end{equation}
are $2\times 2$ and $6\times2$ matrices, respectively; and in $n$ 
we have used vector notation to emphasize that $\vec{c}$, 
$\vec{d}$ and $\vec{0}$ are $3\times 1$ columns.  Of the 8 complex 
parameters $\{a,b,\vec{c},\vec{d}\}$, only $a$ is present in the standard 
model (where it corresponds to the right-handed neutrino's 
majorana mass).  The remaining 7 parameters $\{b,\vec{c},\vec{d}\}$ 
present a puzzle -- they are an unwanted blemish that must be 
removed in order to match observations.  Traditionally, they are 
removed by introducing an extra assumption (namely, that $D$ 
commutes with $L_{a}$ for $a=\{\lambda,q_{\lambda},0\}\in A$)
\cite{Chamseddine:2006ep}; 
but, as emphasized by Chamseddine and Connes (see {\it e.g.}\ 
Sec.~5 of Ref.~\cite{Chamseddine:2007hz}), this {\it ad hoc} 
solution is unsatisfying, and cries out for a better understanding.

Our reformulation yields a simple and satisfying solution to 
this puzzle.  We have seen that the associativity of $B=\Omega 
A\oplus H$ implies the usual order zero and order one 
constraints ($[L_{a},R_{b}]=0$ and $[[D,L_{a}],R_{b}]=0$);
but notice that it also implies a new constraint: $[[D,L_{a}],[D,R_{b}]]=0$.  
This constraint may be satisfied in four different ways, by setting
(i) $b=\vec{c}=\vec{d}=0$;  (ii) $y_{q,11}=y_{q,21}=b=0$; 
(iii) $y_{l,11}=y_{l,21}=\vec{c}=\vec{d}=0$;  or (iv) $y_{l,11}=
y_{l,21}=y_{q,11}=y_{q,21}=\vec{c}=0$.  Note, in particular, that 
solution (i) precisely corresponds to setting the 7 unwanted parameters 
($b,\vec{c},\vec{d}$) to zero, without the additional {\it ad hoc} 
assumption described above!

We can go further by noting that the general embedding of 
$\mathbb{C}$ in $\mathbb{H}$ is given by $q_{\lambda}(\hat{n})
={\rm Re}(\lambda)\mathbb{I}_{2\times2}+{\rm Im}(\lambda)\hat{n}
\cdot\vec{\sigma}$, where $\vec{\sigma}$ are the three Pauli matrices, 
and $\hat{n}$ is a unit 3-vector specifying the embedding direction.
Since all of these embeddings are equivalent, the diagonal 
embedding $q_{\lambda}=q_{\lambda}(\hat{z})$ in Eq.~(\ref{L_a}) 
was arbitrary, and may be replaced by the more
general possibility $L_{a}L_{R}=q_{\lambda}(\hat{n}_{l})L_{R}$
and $L_{a}Q_{R}=q_{\lambda}(\hat{n}_{q})Q_{R}$.  If we redo the 
preceding analysis with this modification, the four solutions
for $D$ are modified accordingly: in particular, in solution (i),
$D$ is given by Eq.~(\ref{D}), where the $2\times2$ matrices $y_{l}^{}$ 
and $y_{q}^{}$ are arbitrary, the $6\times2$ matrix $n$ vanishes, and the 
$2\times2$ matrix $m$ is given by $m=P^{T}M P$, with $M$ an 
arbitrary $2\times 2$ symmetric matrix and $P=(\mathbb{I}_{2\times2}
+\hat{n}_{l}\cdot\vec{\sigma})/\sqrt{2}$ a projection operator.  Then,
one can check the following result: given the arbitrary $2\times2$ matrices 
$y_{l}$, $y_{q}$ and $M$, there is a preferred choice for the embedding 
directions $\hat{n}_{l}$ and $\hat{n}_{q}$ such that, after a change of 
basis on $H$, $L_{a}$ is given by Eq.~(\ref{L_a}) [with the diagonal
embedding $q_{\lambda}=q_{\lambda}(\hat{z})$], $D$ is given by 
Eq.~(\ref{D}), $m$ and $n$ are given by Eq.~(\ref{mn}) with 
$b=\vec{c}=\vec{d}=0$, while $y_{l}$ and $y_{q}$ are given by:
\begin{equation}
  y_{l}\!=\!\left(\!\begin{array}{cc}
  y_{\nu}\varphi_{1} & -y_{e}\bar{\varphi}_{2} \\
  y_{\nu}\varphi_{2} & +y_{e}\bar{\varphi}_{1}\end{array}
  \!\right),\qquad
  y_{q}\!=\!\left(\!\begin{array}{cc}
  y_{u}\varphi_{1} & -y_{d}\bar{\varphi}_{2} \\
  y_{u}\varphi_{2} & +y_{d}\bar{\varphi}_{1}\end{array}
  \!\right),
\end{equation} 
with $\{y_{\nu},y_{e},y_{u},y_{d},\varphi_{1},\varphi_{2}\}\in\mathbb{C}$.
This is precisely the finite geometry that (after fluctuation and substitution into
the spectral action) generates the standard model of particle physics
(see Refs.~\cite{Chamseddine:2006ep, Chamseddine:2007hz, 
Chamseddine:2007ia, vandenDungen:2012ky})!  

This strikingly successful match between the geometric structure on the 
one hand, and the standard model Lagrangian on the other, appears to
provide significant additional evidence: (i) for the suitability 
of Connes' NCG framework for describing the standard model, and
(ii) for the appropriateness of the reformulation presented here.

\section{Discussion}

We end with a few brief remarks.  (i)  In this paper, although we have developed 
a formalism suited to non-associative geometry, our main application has 
been to the associative finite geometry that describes the standard model of 
particle physics.  In a forthcoming paper \cite{NonAssocExamples} we present 
a family of geometries (and their associated spectral actions)
that provide a nice illustration of our formalism in the fully non-associative case.
(ii) Refs.~\cite{Chamseddine:2007hz, Chamseddine:2007ia} observe that
the standard model algebra $A=\mathbb{C}\oplus\mathbb{H}
\oplus M_{3}(\mathbb{C})$ analyzed above can be understood more 
deeply as a subalgebra of $A'=M_{2}(\mathbb{H})\oplus M_{4}(\mathbb{C})$
(see also \cite{Devastato:2013oqa}).  What new light does our formalism shed on this observation?  
(iii)  We have seen that the finite geometry $T$ that encodes the standard model corresponds
to an algebra $B$ that is associative.  But, to evaluate the spectral action, one then tensors
this finite geometry with a continuous geometry to form a new geometry $T'$, and one
can check that the corresponding algebra $B'$ is {\it not} associative
(when one goes beyond the order one associators).  In this sense, non-associativity 
already appears in the traditional NCG embedding of the standard model.  
It is interesting to consider whether this non-associativity might be connected to the 
generalized inner fluctuations considered in \cite{Chamseddine:2013sia, 
Chamseddine:2013rta}, which bear a striking resemblance to the inner derivations 
of a non-associative (and in particular, an alternative) algebra \cite{Farnsworth:2013nza}.  
(iv) In a follow-up paper \cite{Farnsworth:2014vva}, we have shown how the formalism
presented here yields a new perspective on the symmetries of a non-commutative
geometry and, in particular, when we apply this new understanding to the spectral triple
traditionally used to describe the standard model ({\it i.e.}\ the spectral triple 
discussed in the previous section), we find that it actually predicts a slight extension
of the standard model, with two new particles: a $U(1)_{B-L}^{}$ gauge boson, and 
a complex scalar field which carries $B-L$ charge and is responsible for ``higgsing"
the new $U(1)_{B-L}$ gauge symmetry.  These two new particles have important 
phenomenological and cosmological consequences that will be analyzed in 
subsequent work.  

{\bf Acknowledgements.}  Part of this work was carried out at the ``Noncommutative Geometry and 
Particle Physics" Workshop at the Lorentz Center, in Leiden; we thank the
organizers and participants, and particularly Ali Chamseddine and 
Alain Connes for helpful input.  Research at the Perimeter Institute is 
supported by the Government of Canada through Industry Canada and 
by the Province of Ontario through the Ministry of Research \& Innovation.
LB also acknowledges support from an NSERC Discovery Grant.

\end{document}